\documentclass{emulateapj}

\begin{document}

\shorttitle{Parameters and Predictions for HD~17156b}
\shortauthors{Irwin et al.}

\title{Parameters and Predictions for the Long-Period Transiting Planet HD~17156b}

\author{Jonathan Irwin, David Charbonneau\altaffilmark{1} and Philip Nutzman}
\affil{Harvard-Smithsonian Center for Astrophysics, 60 Garden St., Cambridge, MA 02138}
\email{jirwin at cfa dot harvard dot edu}

\author{William F. Welsh and Abhijith Rajan}
\affil{Dept of Astronomy, San Diego State University, 5500 Campanile Dr., San Diego, CA 92182}

\author{Marton Hidas, Timothy M. Brown and Timothy A. Lister}
\affil{Las Cumbres Observatory, 6740 Cortona Dr. Suite 102, Santa Barbara, CA 93117}

\author{Donald Davies}
\affil{23819 Ladeene Ave., Torrance, CA 90505}

\author{Gregory Laughlin and Jonathan Langton}
\affil{UCO/Lick Observatory, University of California at Santa Cruz, Santa Cruz, CA 95064}

\altaffiltext{1}{Alfred P. Sloan Research Fellow}

\begin{abstract}
We report high-cadence time-series photometry of the recently-discovered
transiting exoplanet system HD~17156, spanning the time of transit on
UT 2007 October 1, from three separate observatories.  We present a
joint analysis of our photometry, previously published radial velocity
measurements, and times of transit center for 3 additional events.
Adopting the spectroscopically-determined values and uncertainties for
the stellar mass and radius, we estimate a planet radius of $R_{p} =
1.01 \pm 0.09 \, R_{\rm Jup}$ and an inclination of $i =
86.5^{+1.1}_{-0.7}$~degrees.  We find a time of transit center of
$T_{c} = 2454374.8338 \pm 0.0020$~HJD and an orbital period of $P =
21.21691 \pm 0.00071$~days, and note that the 4 transits reported to
date show no sign of timing variations that would indicate the
presence of a third body in the system.  Our results do not preclude
the existence of a secondary eclipse, but imply there is only a 9.2\%
chance for this to be present, and an even lower probability (6.9\%)
that the secondary eclipse would be a non-grazing event.  Due to its
eccentric orbit and long period, HD~17156b is a fascinating object for
the study of the dynamics of exoplanet atmospheres.  To aid such
future studies, we present theoretical light curves for the variable
infrared emission from the visible hemisphere of the planet throughout
its orbit.
\end{abstract}


\section{Introduction}
\label{intro_sect}

Twenty-eight transiting planets are now reported in the
literature\footnote{See {\tt
http://www.inscience.ch/transits/}}, and the doubling time
scale for new detections is now roughly one year.  As reviewed by
\citet{char2007}, it is these objects that have allowed the study of
the physical structure and atmospheric chemistry and dynamics of gas
giant exoplanets, and opened the field of comparative exoplanetology.
However, the majority have orbital periods of a few days, due both to
the lower geometric probability of transits occurring as the orbital
period increases, and due to the limitations of ground-based
photometric transit surveys that strongly favor detection of
short-period systems.  The latter problem can be circumvented by
searching for transits of known radial velocity planet-bearing stars,
putting to use the radial velocity information to constrain the
possible time of transit.  This is the primary aim of the
Transitsearch.org network \citep{sea2003}\footnote{See {\tt
http://www.transitsearch.org}}, which, starting in 2002, has been
conducting photometric follow-up observations of known radial velocity
detected planets.  This network is a prime example of successful
collaboration between amateur and professional astronomers.

Although longer period planets possess a lower geometric probability to
transit, there exists a loop-hole for some planets on highly eccentric
orbits.  The transit probability for a planet on an eccentric orbit,
with periastron near inferior conjunction, is amplified by a factor
\begin{equation}
A = \left[ {{1+e\cos(\frac{\pi}{2} - \omega)} \over{1-e^2}}\right]
\end{equation}
where $e$ is the eccentricity, and $\omega$ is the argument of
pericenter \citep{sea2003}.  An amplification factor near 2.9 for the
planet orbiting HD~17156 \citep{fischer2007}, along with an extremely
short time window for possible transits, brought the system to the
attention of the Transitsearch.org network.  This was rewarded by the
detection of transits as recently announced by \citet{bar2007}.

This unique system contains a $3.1$-$M_{\rm Jup}$ transiting planet in
a $21.2$ day highly-eccentric orbit ($e = 0.67$) about a bright ($V =
8.2$) G0V star.  \citet{gil2007} have recently presented refined
estimates of the system parameters based on new photometric data.
Even among planets with periods beyond 5 days, where the timescale for
tidal circularization quickly grows to exceed the age of most systems,
HD~17156b's eccentricity is unusually high, thus making it an
interesting test case for models of planetary migration.  Preliminary
measurements of the Rossiter-McLaughlin effect
\citep{rossiter24,mclaughlin24,gw2007,winnasp2007} by \cite{nar2007}
show evidence for a 
misalignment between the stellar spin axis and the planetary orbit
axis, which may indicate a migration mechanism involving planet-planet
scattering \citep[see e.g.][]{chat2007}, or Kozai migration under
perturbation by a yet undetected stellar companion or second planet
\citep{fab2007,wu2007}.  The large eccentricity and long period also
make HD~17156 a particularly attractive target for several additional
follow-up studies.  First, it presents a unique opportunity for the
study of the structure and dynamics of a gas-giant atmosphere under
strongly varying illumination, through infrared monitoring of the
planetary emission (e.g. \citealt{har2006}; \citealt{cow2007};
\citealt{knut2007a,knut2007b}).  Second, by monitoring
successive times of transit, the presence of additional bodies in the
system can be detected or constrained from the resulting perturbations
on the orbit of HD~17156b (\citealt{hm2005}; \citealt{agol2005};
\citealt{sa2005}).

The purpose of our paper is three-fold: First, we seek to refine the
estimates of the system parameters that were only poorly constrained
by the light curves of \citet{bar2007}.  Second, we document the time
of center of transit and search for transit timing variations.  Third,
we evaluate the likelihood that a secondary eclipse will be observable
for HD~17156, and present theoretical predictions of the infrared phase
variations as might be detected with the {\it Spitzer Space
Telescope}.  We elected to perform an independent analysis from that
of \citet{gil2007}, and therefore do not use their orbital parameters
as constraints, since at the time of writing these are still subject
to change as their paper has not yet been accepted for publication.
We do however make use of the time of mid-transit presented in their
Table 1.

We begin by presenting a description of our observations
of the transit event on UT 2007 October 1 in \S \ref{obs_sect}.  We
then present our combined analysis of the extant radial velocity
measurements (\S \ref{rv_sect}) and photometric data (\S
\ref{phot_sect}), and our resulting estimates of the system
parameters and likelihood of a secondary eclipse.  In \S
\ref{disc_sect}, we compare our constraints on the planet mass and
radius with planetary structural models, summarize the current
constraints on the presence of transit timing variations in the
system, and present model predictions for the planetary infrared light
curve.  We conclude in \S \ref{summ_sect} with a discussion of
compelling avenues for future research.

\section{Observations}
\label{obs_sect}

Using the orbital period given by \citet{fischer2007} and the time of
mid-transit measured by \citet{bar2007}, we predicted that an event
would occur on UT 2007 October 1 with transit center at UT 7:53, which
was well-situated for observatories in the southwestern United States.
We present below a description of the data we gathered from three such
observatories, and summarize the calibration of the raw frames and
extraction of the time series for each.

\subsection{Mount Laguna Observatory}

We observed HD~17156 with the Smith 0.6~m telescope at Mount Laguna
Observatory using a Bessell $R$ filter and an SBIG STL-1001E CCD
camera (giving a scale of $0\farcs4$/pixel at the $f/20$ Cassegrain
focus), using BD+71~168 ($V = 9.57$, B8 spectral type) as our local
comparison star.  The integration time for each exposure was 4~s, with
a 4.3s readout 
time and a net cadence of 8.3~s.  For each frame, we calculated the
time at mid-exposure and corrected this time to heliocentric Julian
Day.  We dark-subtracted and flat-fielded the raw images in IRAF, and
then performed aperture photometry using a circular aperture with
radius 12.8~arcsec.  The use of the large aperture insulated our
analysis against seeing-induced aperture losses, as the seeing varied
from 2.1~arcsec to 2.7~arcsec over the course of the night.
Clouds were present at several times during the observations,
especially after mid-transit.  To avoid systematic effects due to
cloud-induced differential extinction of the different color stars, we
rejected observations for which the calibration star showed more than
$8\%$ light loss.  The resulting 2148 points were binned 20:1 and
uncertainties were estimated from the RMS scatter in each bin divided
by $\sqrt{20}$.  Typical uncertainties for the binned points were
$0.3-0.7 \%$.

\subsection{Torrance California}

We observed a field centered on HD~17156 using a commercial $0.25$m
Meade LX200 telescope and a Meade Deep Sky Imager Pro II CCD camera.
We employed an f/3.3 focal reducer to increase the field-of-view to
$\sim 20 \times 30$ arcmin.  We gathered $10\,345$ exposures through
an Edmund Optics IR-pass filter (approximating the conventional
$I$-band), between UT 4:20 and 13:25.  The integration time for each
image was 2~s.  Weather conditions were clear.  Approximately $10\%$
of the frames were lost due to an intermittent issue with the CCD
readout, and due to wind shake, and the guide star was lost
temporarily at 6:10 UT, resulting in a large pointing shift ($\sim 5\
{\rm arcmin}$).

These images were processed using the software described in
\citet{i2007a}, which was originally developed for the Monitor project
\citep{aigrain2007}.  Individual frames were processed by subtracting
a master dark and dividing by a master flat field.  We then ran the
pipeline source detection software (see \citealt{i2007a}) on all of
the data frames, choosing a single frame taken in good conditions for
use as a reference.  For each image, we derived an astrometric
solution employing an implementation of the triangle matching
algorithm of \citet{groth86} to cope with shifts in the telescope
position (these were $\sim 10$ pixels peak-to-peak, in a periodic
fashion, presumably due to errors in the telescope worm drives), and
field rotation due to polar misalignment of the mount.

We generated light curves using the method described in
\citet{i2007a}, using the nearby star BD+71 168 as the comparison
source for the differential photometry.  We found that the normalized
time series data gathered before and after the large image motion at
UT 6:10 have an photometric offset of roughly 3\%.  Since the data
obtained before UT 6:10 were well before the ingress, we simply
trimmed them from the time-series; similarly, we trimmed the data that
occurred more than 0.2 days after the transit midpoint.  We also
trimmed single large outliers by identifying all points that deviated
from the median of the time-series by more than 8\%.  The typical
photometric errors in the final, trimmed time-series are 2\% per data
point, as determined from the RMS variation from the data gathered
after egress.  We then binned this time series by a factor of 15 (and
assumed that the errors in the binned data were reduced according to
Poisson statistics); this binning was necessary to keep the
computational time for the model fitting (\S \ref{phot_sect}) to a
manageable duration, but the binning did not degrade the quality of
the fits.

\subsection{Las Cumbres Observatory Global Telescope}

We also observed the event using a $0.4$m Meade RCX400 telescope on a
custom mount, temporarily located in the car park of the Las Cumbres
Observatory Global Telescope (LCOGT) offices in Goleta, California.
We used an SBIG STL-6303E CCD camera to image a $20\times 30$~arcmin
field with 0.6-arcsec pixels.  380 useful images were taken between UT
5:30 and 11:00, with the airmass decreasing from 1.6 to a minimum of
1.25 near UT 10:00.  No autoguiding was used, and the field drifted by
$\sim 100$ arcsec during the run.  The observations were made through
the ``Red'' filter from the LRGBC filter set provided with the camera
by SBIG,\footnote{\tt
http://www.sbig.com/large\_format/filterchart\_large.htm} with $\sim
90$\% transmission in the range 580--680~nm.  Exposure times varied
from 20 to 30 seconds. The readout time was 22 seconds.  The telescope
was defocused to avoid saturating the target star, which was the
brightest in the field.  The amount of defocus was increased at about
UT 06:30, shortly before the transit ingress, but this did not
significantly affect the relative photometry at that point.

Bias and dark subtraction and flat fielding were done using standard
IRAF tasks. HD~17156 and ten reference stars were measured using the
DAOPHOT aperture photometry package, with apertures of radius
10.6~arcsec.  The light curve was divided by the average of the
reference light curves to remove the effects of varying atmospheric
extinction and focus changes. The standard deviation of the (unbinned)
points outside the transit was 0.4\%.

The scatter of these data is lower than for the Mount
Laguna $0.6$m observations, despite the smaller aperture, since the
larger field-of-view permits the use of multiple comparison stars.  In
the Mount Laguna data, the signal-to-noise ratio of the final light
curve was limited by that of the comparison light curve, due to the
use of a comparison star fainter than the target.  By using multiple
comparison stars for the Las Cumbres data, it was possible to attain a
higher signal level in the comparison light curve, such that the
overall noise was dominated by the target, hence giving a net overall
improvement in signal-to-noise ratio despite the use of a smaller
aperture telescope.  Observing conditions were also somewhat worse at
Mount Laguna, and this along with the use of defocus to render the
observations insensitive to seeing variations, and to average out
pixel-to-pixel flat fielding errors, also contribute to reducing
the noise level in the Las Cumbres data.

\section{Analysis of the Radial Velocity Data}
\label{rv_sect}

Our orbital model was obtained by jointly fitting the
\citet{fischer2007} radial velocity data (converted to HJD) in
conjunction with the following four observed central transit times,
numbered $N = -1,0,2,3$ (where we denote our observation on UT 2007
October 1 as event $N = 0$): $T_{c,-1} = {\rm HJD}\, 2454353.61 \pm
0.02$ \citep{bar2007}, $T_{c,0} = {\rm HJD}\,2454374.8338 \pm 0.0020$
(this paper; see \S \ref{phot_sect}), $T_{c,2} = {\rm
  HJD}\,2454417.2645 \pm 0.0021$ \citep{nar2007}, and $T_{c,3} = {\rm
  HJD}\,{2454438.4835}^{+0.0009}_{-0.0025}$ \citep{gil2007}.

For a Keplerian orbit, the instantaneous stellar radial velocity is  
given by
$V_{{\rm mod},i} = K \, [\cos(f_{i} + \omega)+e \, \cos\omega]$,
where $f_{i}$ is the true anomaly of the planet at time $t_i$.
If we assume an edge-on orbit, the time of central transit occurs when
$f_{t} = {\pi/2}-\omega$.
Given the precision of the available photometry, this approximation  
to $f_{t}$ is excellent.
The true anomaly, $f$, is in turn related to the
the eccentric anomaly, $E$, via
$E = 2\arctan{ \sqrt{(1-e)/(1+e)} \tan(f/2) }$,
so that a central transit, $T_{c,i}$, occurs at a fixed interval
$\Delta T = P(E-e\sin{E})/{2\pi}$
following the epoch of a periastron passage, $T_{p,i}$.

Given $N_{RV}$ radial velocity measurements (here, $N_{RV} = 33$) and
$N_{T_c}$ central 
transit measurements, the goodness-of-fit function of a particular
orbital model is calculated as:
\begin{equation}
\chi^{2} = \sum_{i=1}^{N_{RV}}\left({ V_{{\rm mod},i}- V_{{\rm obs},i}  
\over{\sigma_{i}}}\right)^{2}
+ \sum_{i=0}^{N_{Tc}}\left({ T_{p,i} + \Delta T-T_{c,i}
  \over{\sigma_{i}}}\right)^{2}
\end{equation}
We use a Levenberg-Marquardt algorithm to minimize $\chi^{2}$. The
best-fitting orbital parameters are listed in Table \ref{tbl:params}.
To obtain the quoted uncertainties, a simple bootstrap procedure was
used.  An aggregate of 1000 alternate datasets was created by (1)
redrawing the radial velocities with replacement, and (2) sampling
$T_c$ values for the four transits by drawing from the Gaussian
distributions implied by the quoted errors.  Fits to each of these
data sets were obtained, and the resulting distributions of orbital
parameters were used to derive the $68.3\%$ confidence intervals for
each parameter.  In Table \ref{tbl:params}, we have also given in
brackets the standard deviations of these parameter distributions.
The latter are larger than the former due to the existence of outliers
in the bootstrap sample.  These are generated because the relative
importance of each radial velocity data point in constraining the
derived orbital parameters depends on the orbital phase as a result of
the highly eccentric orbit: the points close to periastron have the
largest effect on the fit.  Consequently, when the bootstrapping
procedure exchanges a point close to periastron, the parameters
estimated from this particular realization of the procedure show
sizable variations.  We therefore prefer the $68.3\%$ confidence
intervals, since these are robust to the presence of the outliers,
which are in turn the result of the bootstrap procedure not being
strictly applicable to data such as these which are not independent
and identically-distributed.

\begin{deluxetable}{lc}
\tabletypesize{\normalsize}
\tablecaption{\label{tbl:params} Fitted Parameters for HD~17156.}
\tablewidth{0pt}
\tablenotetext{a}{For
  the quantities $P$, $e$, ${\omega}$, $K$ and $T_p$, $68.3\%$
  confidence intervals are quoted, with the standard deviation from
  bootstrapping in brackets.  The latter are larger due to outliers in
  the bootstrap sample generated by the high sensitivity of the fitted
  parameters to the velocities near periastron passage.  As discussed
  in the text, the $68.3\%$ confidence intervals are preferred.}

\tablehead{
\colhead{Parameter\tablenotemark{a}} & \colhead{Value}
}

\startdata
$P$                  & $21.21691 \pm 0.00071\ (0.00094) \ {\rm d}$         \\
$e$                  & $0.670 \pm 0.006\ (0.020)$           \\
${\omega}$           & $121{\fdg}3 \pm 0{\fdg}9\ (1{\fdg}8)$             \\
$K$                  & $273.8 \pm 3.4\ (16.3) \ {\rm m \, s^{-1}}$            \\
$T_p$                & $2453738.605 \pm 0.024\ (0.036) \ {\rm HJD}$           \\ 
\\ 
$T_c$                & $2454374.8338 \pm 0.0020 \ {\rm HJD}$ \\  
$i$                  & $86{\fdg}5^{+1\fdg1}_{-0{\fdg}7}$ \\
$R_{p}/R_{\star}$    & $0.070 \pm 0.003$           \\
$R_{p}$              & $1.01 \pm 0.09 \ R_{\rm Jup}$           \\
$R_{\star}$          & $1.47 \pm 0.08 \ R_{\Sun} $           \\
$\rho$               & $3.8^{+0.8}_{-1.1}$ g cm$^{-3}$ \\
\\
$f_1$                & $0.9989 \pm 0.0025$          \\
$k_1$                & $-0.0006 \pm 0.0015$          \\
$f_2$                & $1.006 \pm 0.003$           \\
$k_2$                & $0.0049 \pm 0.0022$          \\
$c_1$                & $-0.0024 \pm 0.0005$         \\
$c_2$                & $0.0173 \pm 0.0029$          \\
$c_3$                & $0.204 \pm 0.064$           \\
\enddata

\end{deluxetable}

We note that the addition of the transit timing data provides a very
strong constraint on the orbital period, but has only a minor effect
on the other orbital parameters, which are in excellent agreement with
the solution given by \citet{fischer2007}.

\section{Photometric Analysis}
\label{phot_sect}

We pooled together the three photometric data sets to fit a model
light curve parametrized by the orbital and physical properties of the
HD~17156 system.  Our model incorporates 4 orbital parameters
initially determined through the fit to the RV data: the period $P$,
eccentricity $e$, time of periastron $T_{p}$, and argument of
pericenter $\omega$.  Given this orbit, we calculate a baseline light
curve for a quadratically limb-darkened star (characterized by
the stellar mass $M_{\star}$, radius $R_{\star}$, and 2 limb darkening
coefficients for each photometric band pass) that is being occulted by
a planet ($r_{p} = R_{p}/R_{\star}$) orbiting at an inclination $i$.
To calculate the baseline light curve, we employ the analytic formulae
of \citet{ma2002}, together with the quadratic limb-darkening
coefficients tabulated by \citet{claret2000,claret2004}, for
$T_{\mathrm{eff}}$ = 6000~K, [Fe/H] = $0.2$, $\log{g} = 4.0$ (based on
the stellar parameters given by \citealt{tak2007}).  We adopted
$R$-band coefficients for the Mount Laguna data, SDSS $r$-band
coefficients for the Las Cumbres data, and $I$-band coefficients for
the Torrance California data.  In the latter two cases, the filters
used for the observations only approximate $r$ and $I$, but the
approximation is more than suitable for our purposes (in particular,
the goodness-of-fit statistic described below is negligibly affected
by errors in the assumed bandpass). The baseline flux is then modified
by observational correction factors unique to each dataset (7
additional parameters described below).

Nominally, our model fixes $P$, $e$, $\omega$, 6 limb darkening
coefficients, and $M_{\star}$, although we iteratively update the first
three of these parameters as the RV fit is apprised of the transit
timing resulting from the light curve fit.  Ultimately, the change in
these three parameters from the iterative update process had
negligible effect on the quality of the light curve fit and had
negligible consequences for the stellar and planetary properties
determined by the transit analysis.  Note that although $T_p$ is a
fit-variable in our transit analysis, the value that is really being
constrained by the photometry is the time of central transit, $T_c$,
which is related to a degenerate combination of $P$, $e$, $\omega$,
and $T_p$.  We fix $M_{\star}$ to the value $1.2 M_{\odot}$
\citep{fischer2007}, which was determined by matching spectroscopic
observations to stellar evolution tracks \citep{tak2007}.  Here, the
uncertainty in $M_{\star}$, for which the 95\% confidence interval is
$1.1-1.3\ M_{\odot}$, has only weak impact on the light curve fit.  This is
because, as described below, we adopt an external constraint on the stellar
radius.

\begin{figure}
\epsscale{1.1}
\plotone{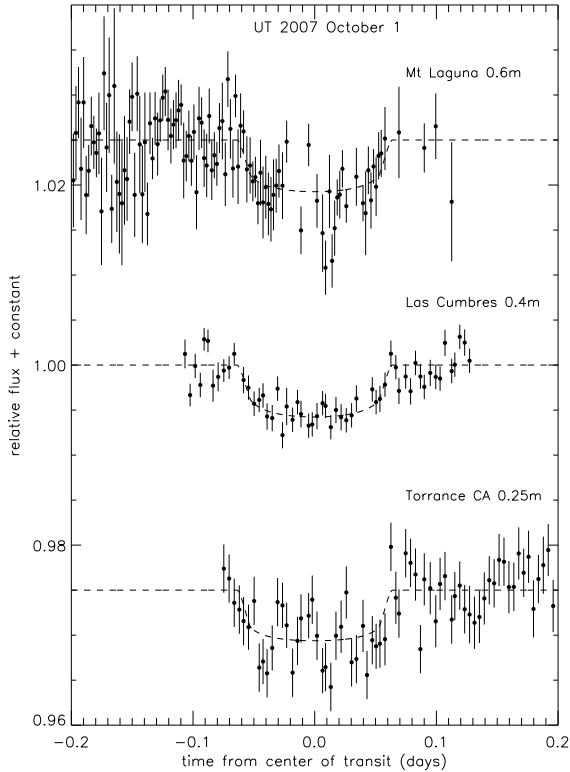}
\caption{Binned light curve of HD~17156, with the airmass and instrumental
corrections described in \S 4 applied. \emph{Top:} Mount Laguna $R$-band
photometry. \emph{Middle:} Las Cumbres photometry in a filter approximating
$r$. \emph{Bottom:} Torrance CA photometry in a filter approximating $I$.
Dashed lines show the model fit simultaneously to the light curve and radial
velocity data, as discussed in the text.}
\label{lc}
\end{figure}

Our light curve model employed 11 free parameters: $T_p$, the
planet-star radius ratio $r_p$, $R_{\star}$, $i$, and 7 additional
parameters related to observational corrections.  The Torrance and
Mount Laguna fluxes were each multiplied by airmass corrections of the
form $f \times \exp(-k a)$, where $f$ is a normalization factor, $a$
is the airmass, and $k$ is the extinction coefficient.  We note, however,
that including the Mount Laguna airmass correction had little effect on the quality of fit (indeed, the determined $f$ and $k$ were consistent with 1 and 0
respectively; see Table \ref{tbl:params}).  We attempted
the same airmass correction for the Las Cumbres data, but found that a quadratic function
of time significantly improved the quality of fit.  We adopted a
$\chi^2$ function as our goodness-of-fit statistic, with an additional
term reflecting a spectroscopic prior on $R_{\star}$, which is
approximately Gaussian with $\sigma = 0.085$ (G. Takeda, personal
communication).  This additional constraint proved necessary, as our
data could not independently determine the stellar radius.  The
goodness-of-fit function is
\begin{equation}
\chi^2 = \sum_{i}{ \left(
\frac{f_{mod}(i)-f_{obs}(i)}{\sigma_i}\right)^2 } + \frac{(R_{\star}/R_{\Sun}-1.47)^2}{0.085^2}
\end{equation}
where $f_{mod}(i)$ is the calculated flux at the time of the $i^{th}$
data point, $f_{obs}(i)$ is the $i^{th}$ flux measurement, and
$\sigma_i$ is the uncertainty of the $i^{th}$ flux measurement
(derived as described for each data-set in \S \ref{obs_sect}).  Using
an IDL implementation of the \texttt{amoeba} algorithm (e.g. see
\citealt{press92}), we performed an initial $\chi^2$ minimization over
the space of free parameters in our model. Using the results, we then
rescaled the $\sigma_i$ so that the reduced $\chi^2$ equals 1
separately for each dataset.  We performed iterations of transit model
fitting in conjunction with radial velocity fitting, using the derived
transit timing to inform the radial velocity fit, which in turn
updated the $P$, $e$, and $\omega$ used for the transit fit.  As
mentioned above, the parameter update had negligible effect on the
derived stellar and planetary properties.  The data, corrected for the
airmass and instrumental effects, are shown in Figure~\ref{lc} along
with the best-fitting solution.

The uncertainty in the transit time was assessed by finely stepping
$T_p$ through values near its best fitting value and calculating the
minimum $\chi^2$ at each step, allowing the other free parameters to
vary.  The 1-$\sigma$ upper and lower limits were assessed by noting
the $T_p$ values at which $\Delta \chi^2 = 1$.  Given the other
orbital parameters, we mapped each $T_p$ into the corresponding time
of central transit, $T_c$.  The resulting best fitting $T_c$ and
uncertainties are reported in Table \ref{tbl:params}.  Over the range
of allowed $T_p$, we found very little correlation between it and the
best fitting stellar/planetary parameters.  For this reason, we decided to
fix $T_p$ at its best fitting value for the Markov Chain Monte Carlo
(MCMC) analysis described below.  Note that by fixing $T_p$, the
MCMC code only needs to perform a full Keplerian orbit calculation
once, rather than at every step in the chain, thus dramatically
speeding up the computation.

\begin{figure}
\epsscale{1.1}
\plotone{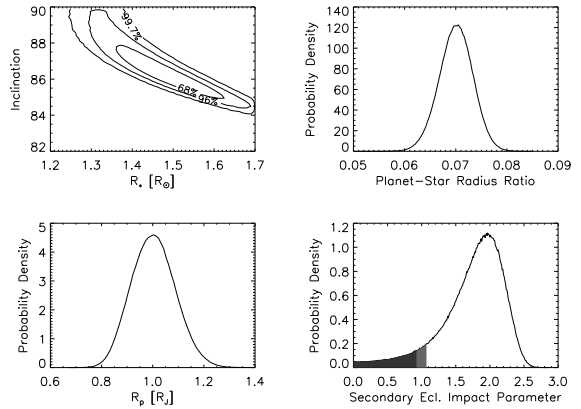}
\caption{Posterior probability distributions for $R_{\star}$, the
planet-star radius ratio, $R_p$, and the secondary eclipse impact
parameter, based on $10^7$ MCMC samples.  The secondary eclipse impact
parameter is the minimum projected separation between planet and star
center (near time of superior conjunction), in units of the
$R_{\star}$.  The dark shaded region shows where full secondary
eclipses occur, while the light shaded region shows where grazing
eclipses occur.}
\label{mcmc}
\end{figure}

To estimate the uncertainty in the 10 remaining model parameters, we
used an MCMC algorithm following the basic recipe described, for
example, in \citet{winn2007}, with whom we share the terminology used
below.  We produced 10 chains, each of $10^6$ points, starting from
independent initial parameter values.  The ``jump function'' was tuned
so that $\sim 25 \%$ of jumps were executed, and the first $10 \%$ of
each chain was discarded to avoid the effect of the initial condition.
We combined the chains, and in Table \ref{tbl:params} we report the
median value of each parameter, and assign uncertainties by taking the
standard deviation of that parameter (except $i$, for which we report
the $68.3\%$ confidence interval).  In Figure~\ref{mcmc}, we show the results
of this MCMC analysis.  One of the major benefits of MCMC is that
probability distributions for derived quantities, such as the
secondary eclipse impact parameter (in the bottom right panel of
Figure~\ref{mcmc}), are produced directly.  Here, the impact parameter was
taken to be the minimum projected separation between planet and star
center in the units of the stellar radius.

\section{Discussion}
\label{disc_sect}

%

\subsection{Planetary mass, radius and density}

The planet's mass of $3.13 \pm 0.11$ M$_{\rm Jup}$ \citep{fischer2007}
is the third most massive among the known transiting planets,
surpassed only by HD~147506b (8.6 $M_{\rm Jup}$; also known as HAT-P-2b;
\citealt{bakos2007}) and XO-3b (13.2 M$_{\rm Jup}$; \citealt{jk2007}).
We find that the mean density of HD~17156 is $\rho=3.8^{+0.8}_{-1.1}$
g cm$^{-3}$, substantially higher than other transiting extrasolar
planets, again with the exception of the remarkable planet HD~147506b
($\rho$=11.9 g cm$^{-3}$; \citealt{loeil2007}).  HD~17156b begins to
bridge the gap between HD~147506b and the other transiting planets.
 
The presence of a large rocky or metallic core has been suggested to
account for ``hot Jupiters'' with high densities.  For HD~17156b, the
models of \citet{boden2003} predict a radius of 1.11 $R_{\rm Jup}$
given the mass of HD~17156b, which is $\sim 1-\sigma$ larger than the
radius derived from the light curve fitting, or equivalently, the
measured density is higher than the models would predict.  However,
for such a massive planet, this discrepancy is difficult to resolve by
invoking the presence of a solid core, since the dependence of
planetary radius on the presence of a core is predicted to be very
weak in this mass regime (e.g. \citealt{boden2003}; \citealt{bur2007};
\citealt{fortney2007a}).  For example, \citet{boden2003} predict that
the addition of a 30 $M_{\earth}$ core reduces the radius by only
$\sim$ 0.01 $R_{\rm Jup}$ for a $\sim$ 3 $M_{\rm Jup}$ planet.  This
scenario would therefore require an extremely massive core to account
for the measured radius of HD~17156b.

Although a large rocky or metallic core can account for planets
with high densities, the low density (i.e. large radius) planets such
as TrES-4 \citep{mand2007} and WASP-1 \citep{cc2007} seem
to defy explanation.  One possible mechanism for keeping planetary radii
from shrinking during planetary evolution is to pump energy into the
planet via tidal interactions (e.g. \citealt{mard2007}). Such
orbital energy transfer mechanisms depend strongly on the orbital
eccentricity, and HD~17156b represents a good test-case for these
theories, having a significantly eccentric orbit ($e=0.67$).  However,
the radius of HD~17156b is not in any way exceptional when compared to
the existing extrasolar planets with close to circular orbits.  This
leads one to speculate that tidal energy transfer via orbital dynamics
may not play a substantial role in planetary radius evolution, or that
in massive planets such as HD~17156b a mechanism is acting to allow
the planets to more rapidly evolve to smaller radii.  We note,
however, that the recently-announced massive planet XO-3b
\citep{jk2007} also has an eccentric orbit, but in contrast to
HD~17156b, its radius is very large ($1.92 R_{\rm Jup}$).  Thus the
challenges posed by newly-discovered exoplanets to theoretical models
of their physical structure seem to continue unabated.

\subsection{Search for transit timing variations}

The availability of multiple measured times-of-transit allows the
presence of additional bodies in the HD~17156 system to be detected or 
constrained, by searching for the influence of their gravitational
perturbations on the orbit of HD~17156b (\citealt{hm2005};
\citealt{agol2005}; \citealt{sa2005}).

\begin{figure}
\epsscale{1.1}
\plotone{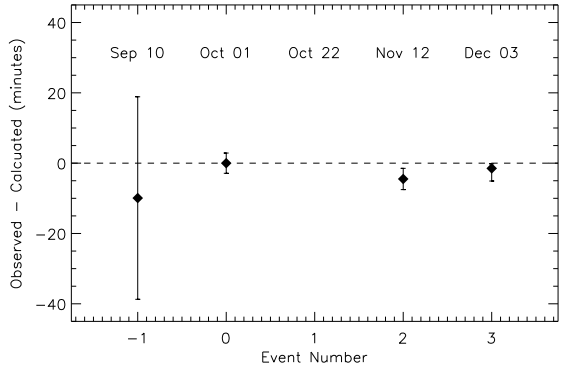}
\caption{Observed minus calculated times of transit for the four
  transit events discussed in this work.  The dashed line shows the
  prediction assuming the values of $T_c$ and $P$ from Table
  \ref{tbl:params}.  Since the value of $P$ was determined from a
  simultaneous fit to the transit times and radial velocity data, it
  does not represent the best fit to the transit times alone.}
\label{oc}
\end{figure}

In Figure~\ref{oc}, we plot the observed minus calculated times of center of
transit for the 4 events reported in the literature, including the
event described in this paper.  We find that these results do not yet
reveal any evidence for timing variations that would indicate the
presence of a third body in the system.  Due to the long orbital
period, HD~17156 may be more amenable to a search for such variations.
However, the large orbital eccentricity implies a large region for
which dynamical stability considerations would exclude the presence of
such planets.

\subsection{Prospects for infra-red observations}

Our photometric analysis directly addresses the probability that
HD~17156b undergoes secondary eclipse, which is only possible for
certain combinations of $R_{\star}$ and $i$.  We find a 9.2\% chance
for secondary eclipses to occur, but only a 6.9 \% chance for
non-grazing eclipses.  If, indeed, HD~17156b undergoes eclipses, $i$
becomes locked rather tightly, thus greatly diminishing the allowed
volume of parameter space for the stellar and planetary properties.
Under this constraint, we find that the new best fitting values for
($i$,$R_{\star}$,$R_p$) are
($88{\fdg}7~^{+1{\fdg}0}_{-0{\fdg}1}$, $1.34 \pm
0.03~R_{\odot}$, $0.89 \pm 0.04~R_{\mathrm{Jup}}$).  These
considerations serve to motivate a search for the secondary eclipses.

HD~17156b's large eccentricity leads to a 25-fold variation in
received stellar flux over the course of the $21.2169\,{\rm d}$
orbital period. This dramatic variation in illumination should produce
complex weather at the planet's photosphere.  In addition, the planet
is subject to strong tidal forces during its periastron passage.  The
planet is therefore expected to be in pseudo-synchronous rotation, in
which the planetary spin will be roughly synchronous with the orbit
during the interval surrounding close approach.  The theory of Hut
(1981) predicts a pseudo-synchronous spin period, $P_ {\rm spin}$ given
by
\begin{equation}
P_{\rm spin}={
1+{15\over{2}}e^{2}+{45\over{8}}e^{4}+{5\over{16}}e^{6}
\over{(1+3e^2+{3\over{8}}e^{4})(1-e^{2})^{3/2}}}P_{\rm orbit}=91.3\, 
{\rm hr}
\end{equation}
Knowledge of $P_{\rm spin}$, $M_{\rm p}$, $R_{\rm p}$, and the
time-dependent pattern of received stellar flux make it possible to
compute global climate models for the planet (e.g. \citealt{sg2002};
\citealt{cho2003}; \citealt{burkert2005}; \citealt{cs2005,cs2006};
\citealt{ll2007}; \citealt{fortney2007b}; \citealt{ddl2007}).  These
models, in turn, can be used to obtain predictions of the planet's
infrared photometric light curve at various wavelengths.  Model light
curves can then be compared with observations using the {\it Spitzer
  Space Telescope} (see, e.g., \citealt{char2005};
\citealt{deming2005}; \citealt{har2006};
\citealt{knut2007a,knut2007b}; \citealt{cow2007}).

We adopt the climate model of \citet{ll2007} and apply it to
HD~17156b, using the known orbital and physical properties of the
planet.  The model uses a compressible 2D hydrodynamical solver
\citep{ada1999} with a one-layer, two-frequency radiative transfer
scheme.  It is important to emphasize that our hydrodynamical model
(like those of other workers in the field) is undergoing rapid
development. Indeed, a primary goal for obtaining physical
observations of planets under time-varying irradiation conditions is
to provide physical guidance to the numerical codes.

\begin{figure}
\epsscale{1.1}
\plotone{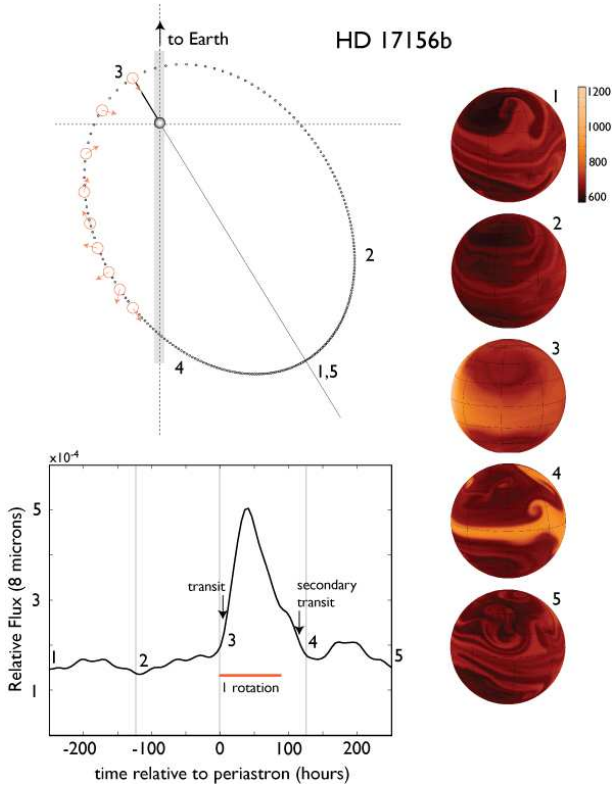}
\caption{{\it Upper left hand panel:} The orbital geometry of
HD~17156b. The small dots show the position of the planet at 2.4 hour
intervals throughout the 21.217~day orbit, and show the planet and
star drawn to scale  ($R_{\star}=1.2 \, R_{\odot}$). The locations
of the planet at successive time intervals equal to one quarter of an
orbit (1-5) are marked. The small orange spheres indicate the 91.3~hour 
predicted spin frequency of the planet, which is expected to be
in pseudo synchronous rotation. 
{\it Right hand panels:} Temperature maps of the planet at locations 1-5,
as viewed from the Earth. The temperature scale is mapped to a black body
color map that approximates the planet's visual appearance in its own
intrinsic radiation. 
{\it Lower left panel:} Predicted $8~{\mu}$m light
curve for the planet during the course of one planetary orbit.}
\label{ir}
\end{figure}

Our model predictions are shown in Figure~\ref{ir}, which, for reference,
includes a model of the planet, the orbit, and the star plotted to
scale (top left). The hydrodynamical model assumes a solar-composition H--He
atmosphere for the planet, and is run for several orbits to achieve
equilibrium, at which point we track the predicted $8\mu$m light curve
for a full orbital period starting at apastron, by assuming a black
body spectrum for each surface element in the model, and integrating
the resulting $8\mu$m flux over the visible hemisphere of the planet.
The right column of Figure~\ref{ir} shows a series of five global
temperature plots, each corresponding to the thermal appearance of the
planet from the Earth.  The figures are equally spaced in time by
one-quarter of an orbit (127.3 hours).

The predicted 8$\mu$m flux from the planet is shown in the lower-left
corner of the figure, and features a rapid rise from a baseline of
$F_{\rm p}/ F_{\star} \sim 1.7\times 10^{-4}$ to $F_{\rm p}/F_{\star}
\sim 5\times 10^{-4}$ during the $\sim 30$ hour interval following
periastron. The magnitude of this rise is comparable with (but
slightly less than) that expected for HAT-P-2b and HD~80606b (Langton
\& Laughlin 2007).

\section{Concluding remarks}
\label{summ_sect}

HD~17156 thus represents a highly interesting system for follow-up
studies in the mid infra-red (e.g. using the {\it Spitzer Space
  Telescope}), even if it does not undergo secondary eclipses.
Ground-based observations should also be pursued to continue the
search for transit timing variations, and hence to constrain the
presence of additional bodies in the system, which, given the high
orbital eccentricity of the planet HD~17156b, may show interesting
dynamical properties and place constraints on planetary formation and
evolution scenarios.  This discovery represents an important success
for the Transitsearch.org project, and highlights the potential
rewards of collaboration between distributed networks of amateur
astronomers and the professional community.

\acknowledgments We thank Mauro Barbieri and the transit discovery
team for alerting us to the October transit opportunity, and Genya
Takeda for discussions regarding the spectroscopic determination of
stellar properties.  DC gratefully acknowledges funding from the David
and Lucile Packard Foundation Fellowship for Science and Engineering.
WFW acknowledges support from Research Corporation.  Observations
obtained at MLO made use of the High Performance Wireless Research and
Education Network sponsored by the NSF ANIR division under grant
ANI-0087344 and the University of California San Diego.  This research
has made use of NASA's Astrophysics Data System (ADS), the SIMBAD
database, operated at CDS, Strasbourg, France, NASA's SkyView and
Dr.~John Thorstensen's SkyCalc software.

Finally, we thank the anonymous referee for a thorough and detailed
report which has helped to improve the paper.

\end{document}